\documentclass[twocolumn,english]{revtex4-1}
\usepackage[T1]{fontenc}
\usepackage[latin9]{inputenc}
\usepackage{geometry}
\usepackage{amsmath}
\usepackage{graphicx}

\makeatletter
\@ifundefined{date}{}{\date{}}
\makeatother

\usepackage{babel}
\begin{document}
\title{Training precise stress patterns}
\author{Daniel Hexner}
\affiliation{Faculty of Mechanical Engineering, Technion, 320000 Haifa, Israel}
\begin{abstract}
We introduce a training rule that enables a network composed of springs
and dashpots to learn precise stress patterns. Our goal is to control
the tensions on a fraction of ``target'' bonds, which are chosen
randomly. The system is trained by applying stresses to the target
bonds, causing the remaining bonds, which act as the learning degrees
of freedom, to evolve. Different criteria for selecting the target
bonds affects whether frustration is present. When there is at most
a single target bond per node the error converges to computer precision.
Additional targets on a single node may lead to slow convergence and
failure. Nonetheless, training is successful even when approaching
the limit predicted by the Maxwell Calladine theorem. We demonstrate
the generality of these ideas by considering dashpots with yield stresses.
We show that training converges, albeit with a slower, power-law decay
of the error. Furthermore, dashpots with yielding stresses prevent
the system from relaxing after training, enabling to encode permanent
memories.
\end{abstract}
\maketitle

\subsection{Introduction}

The macroscopic behavior of materials and physical systems often emerges
from the interactions of a large number of microscopic degrees of
freedom. Examples include, elasticity\citep{goodrich2015principle},
flow patterns in resistor network\citep{rocks2019limits,bhattacharyya2022memory},
and assembly of structures\citep{whitesides2002self}. Engineering
desired behaviors in macroscopic systems is a challanging task due
to the enormous number of coupled degrees of freedom. Recently, there
have been efforts to employ ideas of self-organization, where a material
acquires desired behaviors through training rather than design and
fabrication \citep{pashine2019directed,hexner2020periodic}. As the
system is driven with the training fields the microscopic state evolves
autonomously through inherent plasticity, in a process akin to learning
\citep{scellier2017equilibrium,stern2020supervised2,kendall2020training,stern2021supervised,anisetti2022learning,dillavou2022demonstration}. 

A central challenge in training various behaviors is devising training
rules that exploit the inherent plasticity. To date, a number of training
rules have been proposed, which have allowed to precise strain responses
in elastic systems\citep{hexner2020periodic,stern2020supervised2},
ultra-stable states\citep{hagh2022transient}, and voltage or current
responses in resistor networks\citep{scellier2017equilibrium,stern2021supervised}.
While a precise formalism for learning rules has been proposed\citep{scellier2017equilibrium,stern2020supervised2,kendall2020training,stern2021supervised,anisetti2022learning}
they do not always conform with the microscopic physical laws.

In this paper, we introduce a training rule aimed at manipulating
the stresses in an elastic network\citep{pisanty2021putting,sartor2022predicting}.
We consider a disordered network that is able to evolve through changes
to the rest lengths. Each bond is taken to be a spring and dashpot
in series, whose length changes in proportion to the tension on that
bond. Our goal is to prescribe tension values to a set of randomly
selected bonds. We show that our training method is successful, allowing
us to control the stresses for a large number of bonds. The maximal
number of bonds that can be controlled is bounded by the Maxwell-Calladine
theorem. Training is successful even when approaching this threshold.
We also explore different ensembles for selecting the target bonds,
and show that certain local choices can lead to frustration.

Since dashpots evolve when any tension is present the system cannot
retain the trained stress patterns. To allow permanent memories we
also consider dashpots with a yielding stress. That is, the rest length
evolve only when the tension exceeds a threshold value. We show that,
indeed, this allows to encode permanent memories. Convergence, in
this case, is slow, and is characterized by the power-law decay of
the error. In summary, our work presents a new training rule that
allows to encode arbitrary stress patterns in a model of generic solid. 

\subsection{Training goal \& constraints}

We consider a disordered network of springs which is derived from
a packing of repulsive spheres at zero temperature\citep{Ohern}.
The regime that is considered is far from the jamming transition ($\Delta Z\approx1.5$)
and therefore we expect that this ensemble represents a generic disordered
material\footnote{Previous studies on training found that the particular ensemble is
unimportant. }. The excess coordination number is defined by, $\Delta Z=Z-Z_{c}$,
where $Z=\frac{2N_{b}}{N}$ is the average coordination number (twice
the number of bonds per nodes) and $Z_{c}\approx2d$ marks the minimal
coordination number needed for rigidity. 

\begin{figure}
\includegraphics[scale=0.4]{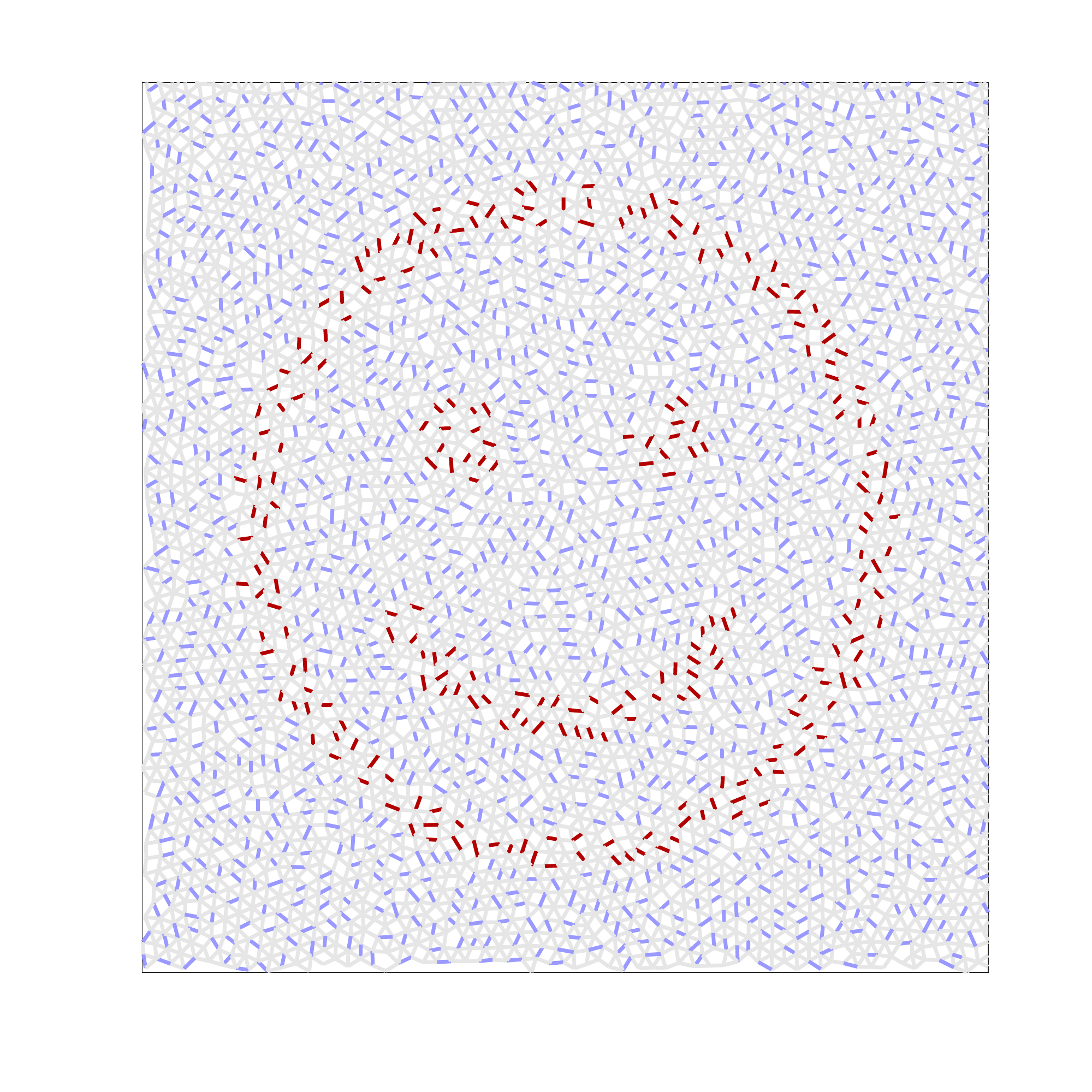}

\caption{An example of training a ``smiley'' pattern. The red bonds are targets
that are trained to have a stress, $t_{0}=10^{-3}$, blue bonds are
trained to have a negative stress, $-t_{0}$ and the remaining grey
bonds are the learning degrees of freedom. Here, the error is approximately
$10^{-7}$. \label{fig:smiley}}
\end{figure}

Our goal is to control the stresses on a set of preselected target
bonds that are chosen randomly. Each of the target bonds is assigned
a desired tension, $t_{T}^{D}=\pm t_{0}$, with equal probability.
In Fig. \ref{fig:smiley} we show an example of a pattern that was
trained to a near perfect response (here, the pattern is non-random
- a smiley face). The target bonds with a tension of $t_{0}$ are
marked in red, while the targets with the tension $-t_{0}$ are ma
in blue. The remaining bonds are the ``learning degrees of freedom''
and their stresses are not indicated. 

Before turning to the training protocol, we first discuss physical
constraints on the patterns a network can acquire. 

\noindent \textbf{Local constraints:} A given node, with $z$ bonds,
is in force balance when the sum of all forces equates to zero:

\begin{equation}
\sum_{b=1}^{z}t_{b}\hat{r_{b}}=0.
\end{equation}
Here, $t_{b}$ are the tensions and $\hat{r_{b}}$ are the unit vectors
of the bonds. When the tensions are small, the bond angles are fixed
and the only degrees of freedom are the tensions on the bonds. Since
there are $d$ equations for force balance the number of independent
degrees of freedom are $z-d$. If the unit vectors are parallel then
the number of independent degrees of freedom is reduced. While perfectly
aligned bonds are non-generic, nearly parallel bonds can be found
with a finite probability, which later we will argue leads to difficulties
during training.

\noindent \textbf{Global constraints:} The tensions on the bonds in
force balance, can decomposed in to a basis of states of self-stress\citep{Lubensky_rev}.
Recall, that the states self stress are the set of independent stresses
that maintain force balance. The number of states of self-stress,
$N_{sss},$ is related to the number of zero modes, $N_{z}$, the
number of nodes, $N$, and the number of bonds, $N_{b}$ through the
Maxwell-Calladine theorem\citep{maxwell1864calculation,CALLADINE,Lubensky_rev}:

\begin{equation}
N_{sss}-N_{z}=N_{b}-dN.\label{eq:Maxwell_Calladine}
\end{equation}
Since the networks are rigid there are no zero modes, except for those
that are contributed from the boundary conditions (translations and
rotations). We neglect these in our count because their number is
small with respect to the number of targets. Assuming the states of
self-stress are extended, the maximal number of independent tensions
that can be set is $N_{sss}$. We denote the number of target bonds
per nodes as $\Delta=\frac{N_{T}}{N}$. The bound on the number of
targets that can be trained, $\Delta_{c},$ is given by, 
\begin{equation}
\Delta_{c}=\frac{N_{sss}}{N}\approx\frac{N_{b}}{N}-d=\frac{Z}{2}-d=\frac{\Delta Z}{2}.
\end{equation}
This implies that highly coordinated network have a larger capacity
and therefore we focus on this regime. 

\subsection{Training protocol}

Next we discuss the training protocol. As noted, there are two populations
of bonds: target bonds, whose tension we aim to control, and the remaining
bonds constitute the learning degrees of freedom; only these bonds
change. We assume that their rest length evolves in proportion to
tension on the bond, similar to a dashpot: 
\begin{equation}
\partial_{t}\ell_{i,0}=\gamma k_{i}\left(\ell_{i}-\ell_{i,0}\right).\label{eq:Evolve}
\end{equation}

Our training protocol consists of alternating between two states.
A \emph{free state} where no external stresses are applied; in this
state we only measure the stresses on the targets, denoted by, $t_{T}^{F}$.
In the free state we also assume (for the time being) that the rest
lengths do not change. We then switch to the \emph{clamped state}
and apply to the targets the clumped stresses, $t_{T}^{C}$. The applied
clamped stresses are different that the desired stresses, $t_{T}^{D}$,
and are chosen to evolve in proportion to the difference between the
desired and free stresses, 

\begin{equation}
\partial_{t}t_{T}^{C}=\lambda\left(t_{T}^{D}-t_{T}^{F}\right).\label{eq:Learn}
\end{equation}
In the clamped state the rest lengths evolve, in accordance to Eq.
\ref{eq:Evolve}, while in the free state we assume the system does
not evolve. In experiment perhaps the free state can be frozen by
either lowering the temperature, or spending very little time in the
free state, so that the system changes very small.

The intuition behind this training rule is as follows. Imagine we
begin in an unstressed state and squeeze a target bond to a desired
$t_{T}^{D}$, while allowing the remaining bonds to evolve according
to Eq. \ref{eq:Evolve}. Eventually, this causes the bonds to evolve
to a state with no internal stress except on the squeezed target bond.
When the targets are unclumped the system cannot return to the initial
unstressed state. The stress on the target bond in the free state
is then given by $t_{T}^{F}=\alpha t_{T}^{D}$, where $\alpha<1$.
To achieve the desired stress we need to increase the value of the
applied clumped stress, $t_{T}^{C}>t_{T}^{D}$. In essence, this protocol
can be considered a control loop where the training signal integrates
over the error. 

We note that $\gamma$ defines a relaxation rate of the internal stresses.
In our simulations we do not wait until the system completely relaxes
in the clamped state\citep{stern2022physical}. At each time step
we compute the stresses in the free and clumped state by in force
balance (by minimizing the energy), and evolve the rest length and
clumped stresses in a single iteration of Eq. \ref{eq:Evolve} and
\ref{eq:Learn}. Note that there are many choices for parameters which
appear to have a weak effect on our results. In the data we present
the parameters are as follows. The system is two dimensional with,
$N=200$, $\lambda\Delta t=0.1$, $\gamma\Delta t=0.1$ (here, $\Delta t$
is the time step), and all the spring constants are identical taken
to be $k_{i}=1.0$. 

\subsection{Results}

Next, we study numerically the success of our training rule. We define
an error which sums over the all the target bonds and is normalized
by the amplitude of the tension: 

\begin{equation}
\eta=\frac{1}{t_{0}}\sqrt{\frac{1}{N_{T}}\sum_{i=1}^{N_{T}}\left(t_{T,i}^{D}-t_{T,i}^{F}\right)^{2}}.
\end{equation}

We will consider two local selection rules for choosing the targets.
First, we allow each node to have at most one target bond and later
on we will consider having at most two target bonds per node. The
number of targets on a given node affects the ease of training, and
the possible resulting frustration. 

\begin{figure}
\begin{centering}
\includegraphics[scale=0.45]{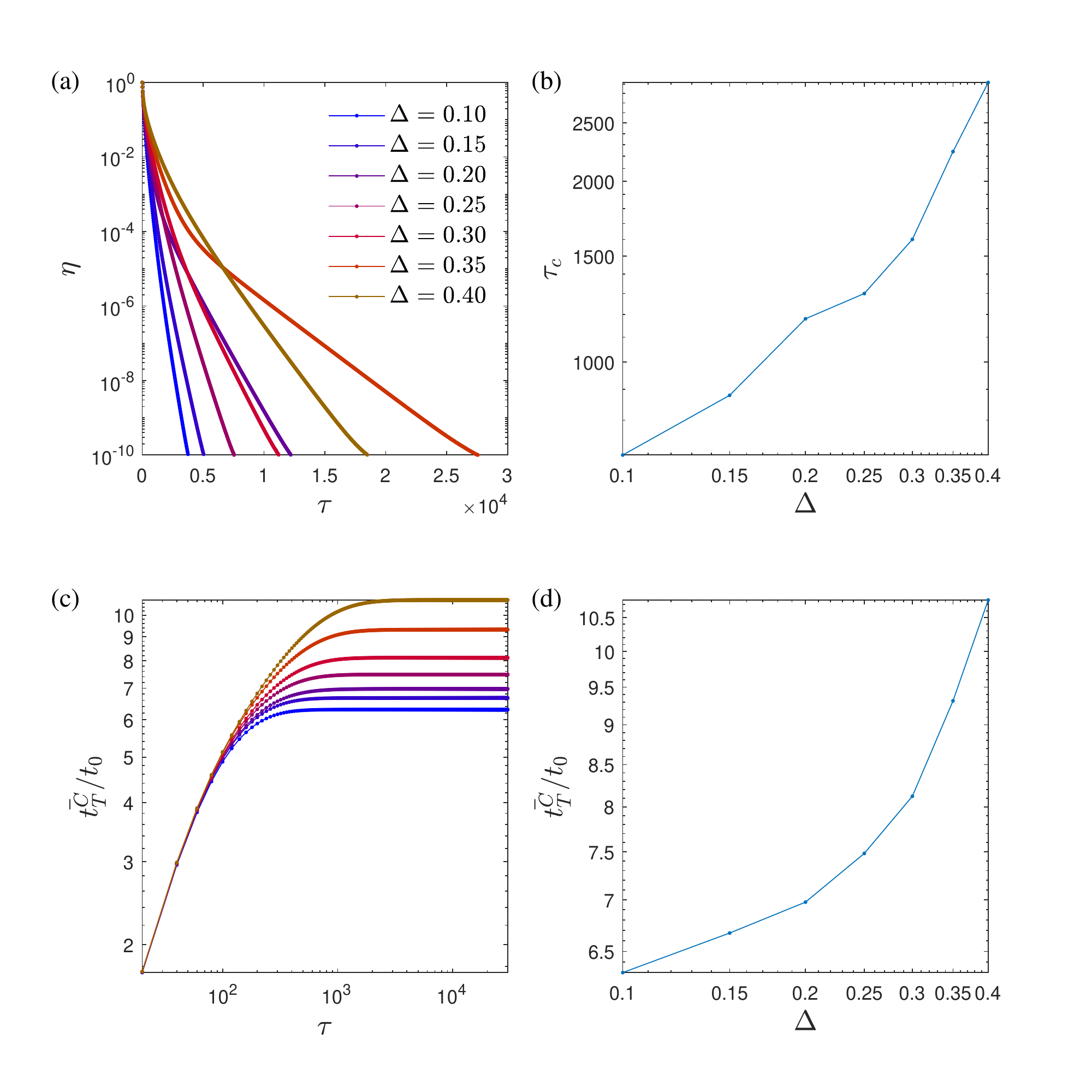}
\par\end{centering}
\caption{Training when each node has at most one target bond. (a) The error
as a function of time for different numbers of targets. (b) The convergence
time, defined by the number of cycles needed to reach $\eta=10^{-3}$,
grows with $\Delta$. (c) The training forces in the clumped state
as a function of time. (d) The training forces in the plateau as a
function of the number of targets. \label{fig:one_per}}

\end{figure}

Fig. \ref{fig:one_per}(a) shows the error as a function of the number
of cycles when each node has at most one target. At large times the
decay is approximately exponential\footnote{In the appendix we compute the convergence rate for a simple case.},
and its decay is limited by computer precision. Increasing the number
of targets yields slower convergence. We estimate the convergence
time by measuring when the the error decreases to a given arbitrary
value, $\eta=10^{-3}$. Fig. (b) shows the convergence time grows
faster than a power-law. This could suggest a divergence at a finite
value of $\Delta=\frac{N_{T}}{N}$, corresponding to a phase transition.
The number of targets are limited by the requirement that each node
has at most one target bond, and therefore we are unable to approach
that transition.

We also consider the strength of the training stresses. Fig. \ref{fig:one_per}(c)
shows the root mean square of the clumped tensions, 
\begin{equation}
\overline{t_{T}^{C}}=\sqrt{\frac{1}{N_{T}}\sum_{i=1}^{N_{T}}\left(t_{T,i}^{C}\right)^{2}}.
\end{equation}
Initially, the training stresses grow with the number of training
cycles and then reach a plateau. The plateau value increase with $\Delta$,
and for the largest considered value, $\overline{t_{T}^{C}}\approx10t_{0}$.
Fig. \ref{fig:one_per}(d) shows that the tension, similar to relaxation
time grows faster than a power-law, again hinting at a possible inaccessible
transition. 

\begin{figure}
\begin{centering}
\includegraphics[scale=0.45]{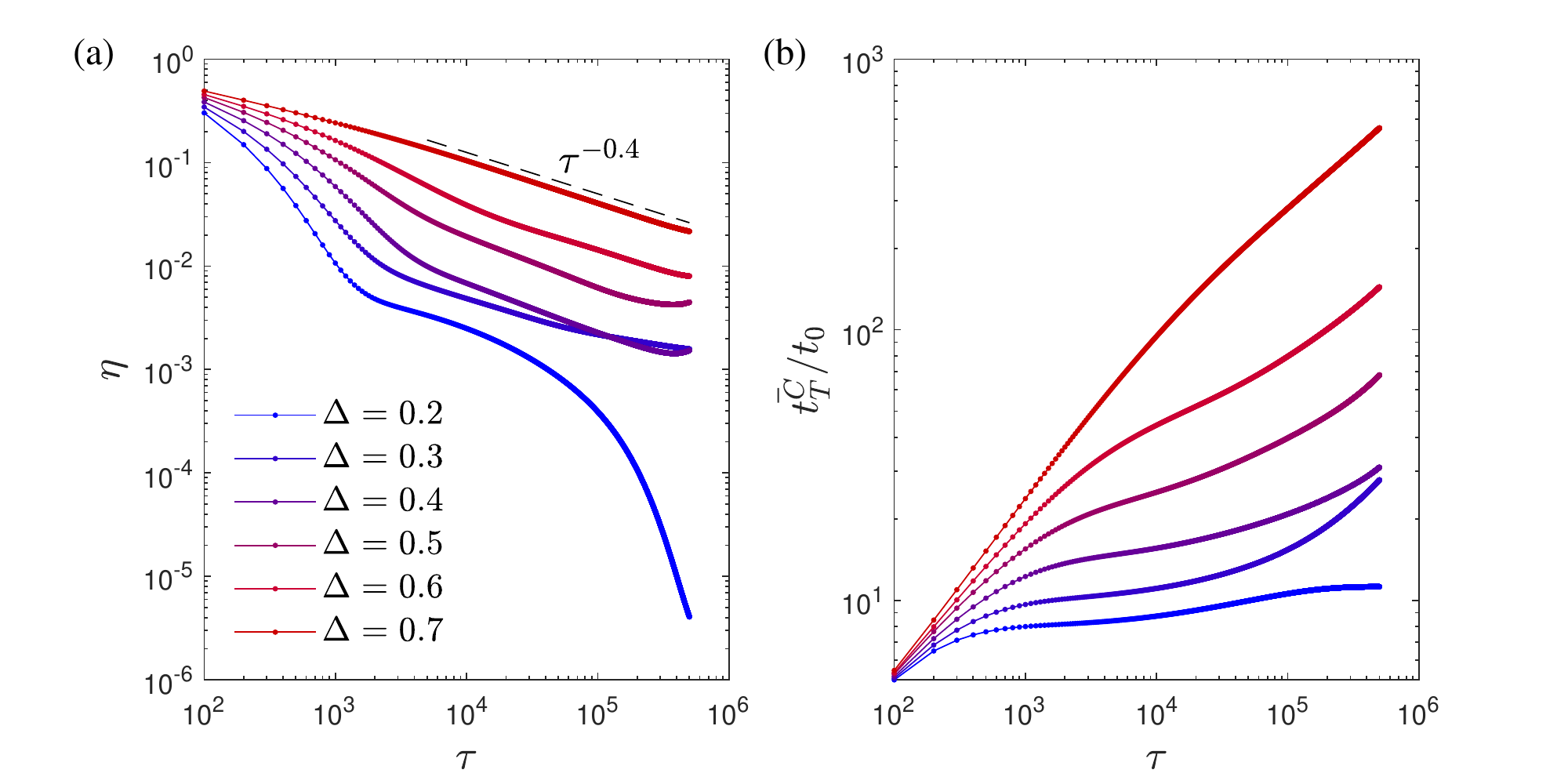}
\par\end{centering}
\caption{Training when each node has at most two target bonds. (a) The training
error as function of time converges slower for large $\Delta$. Note,
that the training error decreases even near the theoretical limit
of $\Delta_{c}\approx0.75$. (b) The training stresses as a function
of time. Here, $t_{0}=10^{-6}.$\label{fig:two_per}}
\end{figure}

Next we consider the effect of allowing two target bonds per node.
Fig. \ref{fig:two_per}(a) shows the error as a function of number
of cycles. As shown, the error decreases even for a large number of
targets close to the maximal capacity, $\Delta_{c}\approx0.75$. Unlike
the case that each node had at most a single target bond, convergence
does not appear to be exponential. A power-law fits reasonably at
intermediate times. The difference in convergence is also expressed
in the applied training tensions, $\overline{t_{T}^{C}}$. Fig \ref{fig:two_per}(b)
shows that the clumped tensions, $\overline{t_{T}^{C}}$, grow monotonically
and do not appear to converge. Continual growth, we argue, is a mechanism
that leads to failure at very large times. Therefore, it preferable
to cease training after a finite number of cycles.

\begin{figure}
\begin{centering}
\includegraphics[scale=0.45]{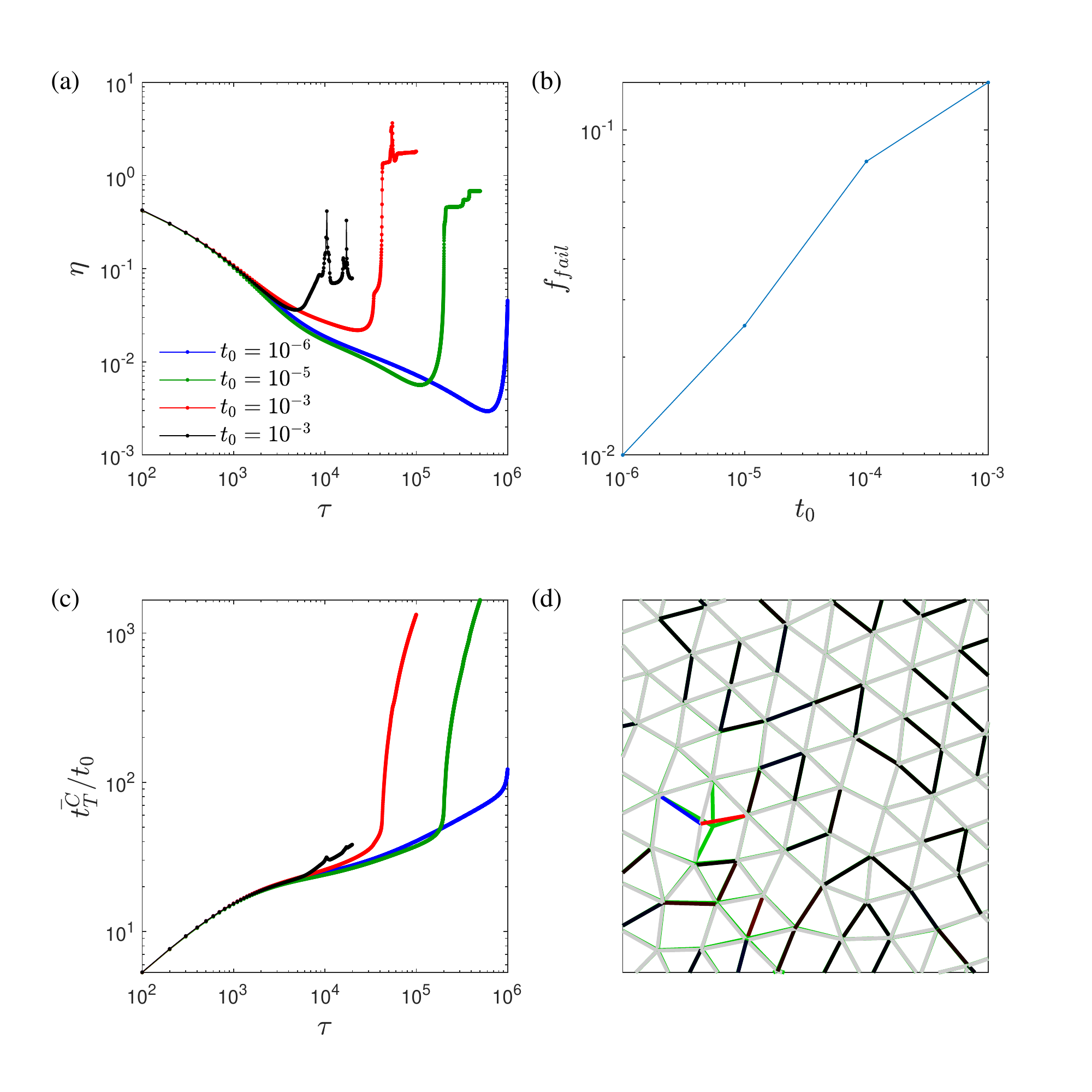}
\par\end{centering}
\caption{Analysis for failure to train when each node has at most two target
bonds. (a) The error as a function of time. (b) The fraction of failed
states as a function of $t_{0}$, define by $\eta>0.1$. (c) The training
stresses as a function of time grow sharply when the error increases.
(d) An example of a realization that failed in the free state. Two
opposing target bonds, where the stress on one bond falls below the
desired (blue) value while on second it is above the desired value
(red). In green, the network in the clumped state differs from the
network in the free state. Here, $t_{0}=10^{-4}.$\label{fig:failure}}
\end{figure}

To study failure we increase $t_{0}$ and characterize how the system
fails. Fig. \ref{fig:failure}(a) shows an example where the error
initially decreases and then begins to steadily increase. The crossover
time between these two regimes decreases upon increasing $t_{0}$.
Namely, training with a small $t_{0}$ is more successful and failure
occurs at much later times. Even though training fails on average,
the majority of realizations are successful. Fig. \ref{fig:failure}(b)
shows the fraction of realizations that failed, defined by $\eta>0.1$.
The fraction of realizations that fail is fairly small, and increases
with $t_{0}$. We also show in Fig. \ref{fig:failure}(c) that failure
is accompanied by a sharp increase in the training tensions. 

To identify how the system fails we have looked at videos of the dynamics.
An example is presented in the supplementary information; here we
present a single frame in Fig. \ref{fig:failure}(d). The learning
degrees of freedom (bonds that are not targets) are plotted in grey.
Targets whose error is small are dark, while blue bonds correspond
to bonds whose tension is below the desired value and red are above
the desired value. In Fig. \ref{fig:failure}(d) we show that the
error is largest at a node where a blue and red bond meet at nearly
$180^{o}$. The two bond have conflicting goals -- one bond would
like to increase the tension while the other would like to decrease
the tension. This conflict causes the training forces on tensions
to continually grow. 

In Fig. \ref{fig:failure}(d) we also show in green the network in
the clumped state. In the regime where $t_{0}$ is small, and when
training is successful the clumped stresses are small. Therefore,
the structure of the clumped state is nearly identical to that of
the free state. However, due to the large forces induced by frustration
the structure of the clumped state becomes very different from the
free state. Thus, the geometry of the clumped state do not represent
that of the free state, which contributes to failure.

In summary, having more than single bond per node may lead to frustration,
which results in the training stresses continually growing and possibly
failure. The frustrated nodes appear to be localized and somewhat
rare but have an overwhelming contribution to the success of training.
This suggests that the physics of rare regions is important\citep{vojta2006rare}.
This picture explains why having a very small $t_{0}$ is beneficial
for training. Small $t_{0}$ allows for the structure in the clumped
state to be similar to that of the free state even when the clumped
tensions are multiples of $t_{0}$. 

\subsection{Permanent memories}

In the previous analysis we relied on the system having two states.
A clumped state where each bond evolves in accordance to the stress
on the bond and the free state which does not evolve. In principle,
if the system is allowed to evolve in the free state all the stresses
will decay to zero. To allow permanent memories, we consider dashpots
with a yielding stress. That is, their rest length changes when the
tension, $t_{i}$, exceeds a threshold value, $t_{th}$, 

\begin{equation}
\partial_{t}\ell_{0,i}=\begin{cases}
\gamma\left(t_{i}-t_{th}\right) & t>t_{th}\\
\gamma\left(t_{i}+t_{th}\right) & t<-t_{th}\\
0 & -t_{th}<t_{i}<t_{th}
\end{cases}.\label{eq:thresh_eq}
\end{equation}
To allow for permanent memories the threshold must exceed the tension
amplitude, $t_{0}$. In our simulations we take $t_{th}=2t_{0}$. 

\begin{figure}
\begin{centering}
\includegraphics[scale=0.45]{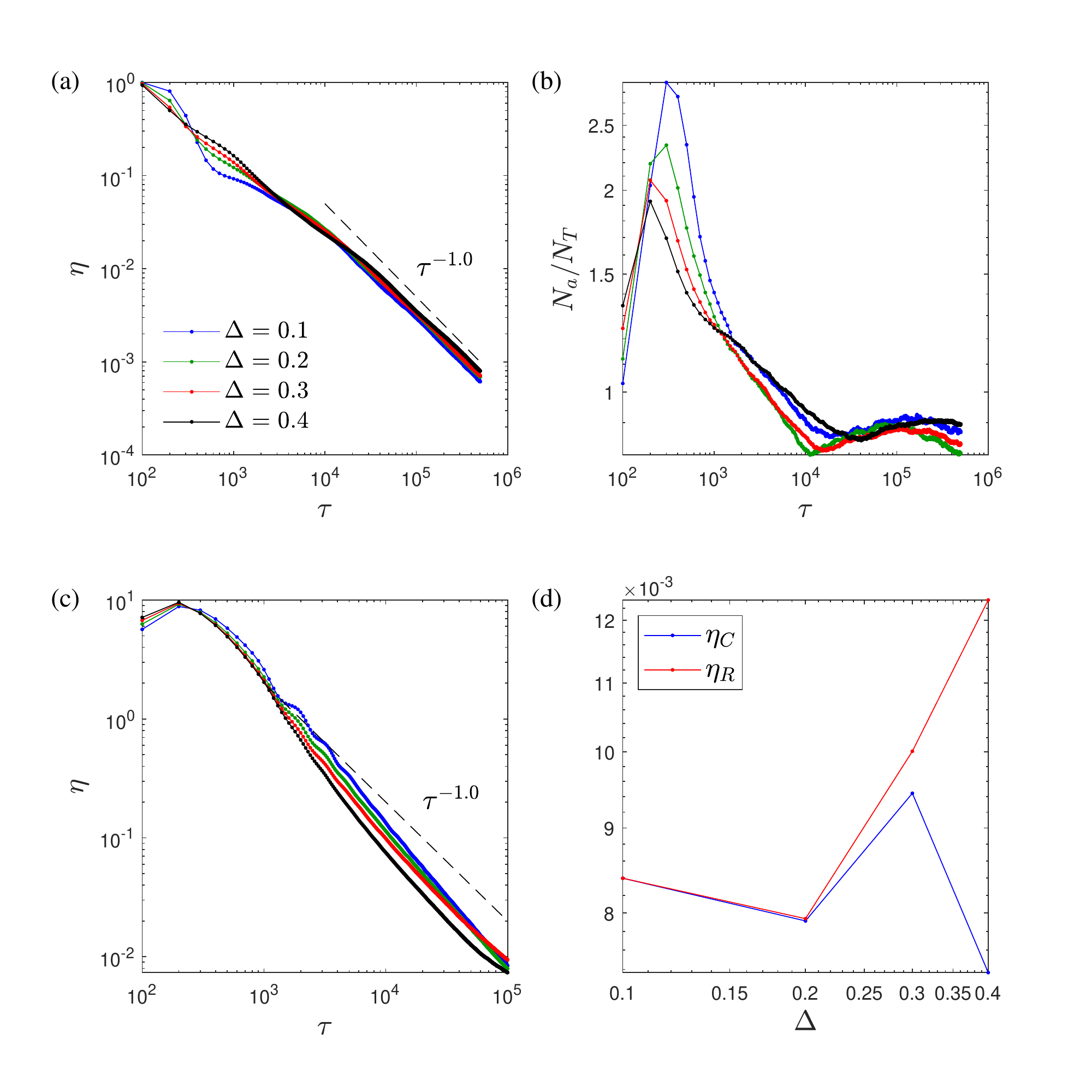}
\par\end{centering}
\caption{Training with dashpots that have a yield stress. (a) The error as
function of time when the network evolves only in the clumped state
is approximately a power-law. (b) The number of bonds that evolve
in a given cycle is of order $N_{T}$. (c) The error as a function
of time when the network evolves in both the clumped and free state
is also approximately a power-law. (d) After the system has been trained
we allow the system to relax in the free state. A comparison of the
error before, $\eta_{C}$, and after the system has completly relaxed
in the free state, $\eta_{R}$. \label{fig:yielding}}

\end{figure}

Here, we avoid frustration effects by allowing each node to have at
most one target site. We first, consider the effect of the yielding
stress when the network does not evolve in the free state. Fig. \ref{fig:yielding}(a)
shows the error as a function of time. Unlike the previous exponential
decay, here, convergence is a power-law, $\eta\propto\tau^{\approx-1.0}$.
The convergence of the error demonstrates the robustness of our results.
We also measure the fraction of bonds that change their length at
a given time, shown in Fig. \ref{fig:yielding}(b). Here, only a small
number of bonds, of order $N_{T}$, are affected by training at any
given time. 

The slow relaxation can be understood through a simple argument. Imagine
that during training the stress on a target bond overshoots beyond
the desired value. To undo this error the applied stress, $t_{T}^{C}$,
must be reversed. However, to reverse the change in the rest length
the applied tensions must be reversed by at least $t_{th}$. Since
the rate of change to $t_{T}^{C}$ is proportional to the error, the
time scale for reversing $t_{T}^{C}$ scales as $\tau\approx t_{c}/\eta$.
Therefore, the error also decreases as $\eta\propto1/\tau$. 

Lastly, we relax our assumption that the free state does not evolve.
We use the same rule, given in Eq. \ref{eq:thresh_eq}, for the free
and clumped state and we allow the system to remain in the free and
clumped state for equal periods of time. Fig. \ref{fig:yielding}(c)
shows the error as a function of time during training. Also here training
is slow to converge. At the end of the training period we then allow
the system to continually evolve in the free state, until all the
tensions fall below $t_{th}$. Fig. \ref{fig:yielding}(d) compares
the error when training is complete and after the system is allowed
to completely relax. After relaxation the error is slightly larger
but overall still small. Since the system has completely relaxed in
a frozen state ($\left|t_{i}\right|<t_{th}$) the trained patterns
are permanent. 

\subsection{Conclusions}

In summary we have shown that a generic network of springs can be
trained to encode a pattern of stresses at randomly selected target
sites. We have introduced a training (learning) rule where the learning
degrees are the rest lengths of the bonds, which evolve in proportion
to the tension on the bond. The difference between the stresses in
the free state and the desired stresses determine the change in training
stresses in the clumped state. This training rule is robust and functions
well even when the dashpots have a yielding stress, which allows to
encode permanent memories. 

We remark that our rule is different than the recent equilibrium propagation\citep{scellier2017equilibrium,stern2020supervised2,kendall2020training,stern2021supervised,anisetti2022learning}
or the coupled learning\citep{stern2021supervised}. Those algorithm
evolve as a function of the difference between the free and clumped
state. In our case, the system evolves only in the clumped state,
though the applied training stresses depend on measurements in the
free state. We argue that our rule is advantageous, since the physical
plasticity law depends only on the current state of the system.

We have also considered the convergence under different scenarios.
We find that convergence depends on the local selection rule of the
targets. When each node has at most a single target, convergence of
the ordinary dashpots is exponential with a time scale that increases
with the number of targets. For the dashpots with yielding stresses,
convergence is approximately a power-law. Having two targets bonds
per node also results in local frustration. Failure may occur when
the training stresses grow without a bounded. 

Our study presents another example of trainable materials. It is interesting
to note that the same physical system of springs and dashpots was
previously used to train strain responses. This demonstrates that
different training rules allow to achieve for the same system different
behaviors. Perhaps our training method could be useful in various
applications, such as, creating optical patterns in photoelastic materials,
designing materials with specific failure modes and possibly useful
in manipulating flow networks. 
\begin{acknowledgments}
I would like to thank Himangsu Bhaumik, Marc Berneman and Yoav Lahini
for enlightening discussions. This work was supported by the Israel
Science Foundation (grant 2385/20) and the Alon Fellowship.
\end{acknowledgments}

\subsection{Appendix \label{subsec:Appendix}}

Here we compute the convergence rate for two springs connected in
series, where one of the springs is the target and the second spring
is the learning degree of freedom. We consider two cases: 1. The system
fully relaxes in the clumped state. That is, at each training cycle
the rest lengths evolve in the clumped state until the tensions on
the bonds vanish. 2. Finite relaxation rate. We assume that the overall
length of the two springs is constrained to be , $L=\ell_{1}+\ell_{2}$,
where $\ell_{i}$ denotes length of the bond. The two spring constants
are taken to be equal, $k$. As above, the rest length of the target
spring, $\ell_{1,0}$ does not change while the rest length of $\ell_{2,0}$
evolves as a function of the stress on that bond in the clumped state.
The superscript $C,F$ denote the clumped and free state respectively.

\subsubsection{Complete relaxation}

We begin by writing the tensions in the free state. 

\begin{align}
t_{1}^{F} & =\frac{k}{2}\left(L-\ell_{1,0}-\ell_{2,0}\right).\label{eq:t1_f}
\end{align}
Next we write the force balance equations in the clumped state: 

\begin{align*}
k\left(\ell_{1}^{C}-\ell_{1,0}\right) & =k\left(\ell_{2}^{C}-\ell_{2,0}\right)+t_{T}^{C},
\end{align*}
where $t_{T}^{C}$ is the training stresses in the clamped state.
Assuming complete relaxation in the clumped state we can set the tension
on bond 2 to zero.

\begin{align*}
k\left(\ell_{1}^{C}-\ell_{1,0}\right) & =t_{T}^{C}\\
\Rightarrow\ell_{2,0}= & \ell_{2}^{C}=L-\ell_{1}^{C}=L-\ell_{1,0}-t_{T}^{C}/k
\end{align*}
We can now express $t_{1}^{F}$ in terms of $t_{T}^{C}:$ 

\begin{align*}
t_{1}^{F} & =\frac{t_{T}^{C}}{2}
\end{align*}
Finally, the training rule evolves the clumped tension in accordance
with, 

\begin{align*}
\frac{d}{dt}t_{T}^{C} & =\lambda\left(t_{0}-t_{1}^{F}\right)\\
 & =\lambda\left(t_{0}-\frac{t_{T}^{C}}{2}\right)
\end{align*}
Convergence is therefore exponential with the rate constant given
by $\lambda/2$, 

\[
t_{T}^{C}=2t_{0}+Ae^{-\frac{\lambda}{2}t}.
\]

\subsubsection{Finite relaxation rate}

Next we allow the rest length to evolve at a finite rate. As before,
we solve the force balance equations, however, now without assuming
that the force on the second bond vanishes. First we solve for $\ell_{2}^{C}$:
\begin{align*}
k\left(\ell_{1}^{C}-\ell_{1,0}\right) & =k\left(\ell_{2}^{C}-\ell_{2,0}\right)+t_{T}^{C}\\
\Rightarrow\ell_{2}^{C} & =\frac{1}{2}\left[-t_{T}^{C}/k+L-\ell_{1,0}+\ell_{2,0}\right]
\end{align*}
The rest length of the second bond evolves via, 

\begin{align}
\frac{d\ell_{2,0}}{dt} & =\gamma k\left(\ell_{2}^{C}-\ell_{2,0}\right)\nonumber \\
 & =\frac{\gamma}{2}\left[-t_{T}^{C}/k+L-\ell_{1,0}-\ell_{2,0}\right]\nonumber \\
 & =\gamma/k\left[-\frac{t_{T}^{C}}{2}+t_{1}^{F}\right]\label{eq:Eq1_appendix}
\end{align}
Note, that in the last step we have plugged in Eq. \ref{eq:t1_f}.
This equation can be recast in terms of $t_{1}^{F}:$

\begin{align*}
\frac{d}{dt}t_{1}^{F} & =-\frac{k}{2}\ell_{2,0}=\frac{\gamma}{2}\left[\frac{t^{C}}{2}-t_{1}^{F}\right]
\end{align*}
The evolution of $t_{T}^{C}$is therefore given by, 

\begin{align}
\frac{d}{dt}t_{T}^{C} & =\lambda\left(t_{0}-t_{1}^{F}\right)\label{eq:Eq2_appendix}
\end{align}
We must solve the two coupled equations Eq. \ref{eq:Eq1_appendix}
and \ref{eq:Eq2_appendix}. To this end we expressed these equation
in matrix form, 

\[
\frac{d}{dt}\left(\begin{array}{c}
t_{1}^{F}\\
t_{T}^{C}
\end{array}\right)=\left(\begin{array}{cc}
-\frac{\gamma}{2} & \frac{\gamma}{4}\\
-\lambda & 0
\end{array}\right)\left(\begin{array}{c}
t_{1}^{F}\\
t_{T}^{C}
\end{array}\right)+\left(\begin{array}{c}
0\\
\lambda t_{0}
\end{array}\right)
\]
To solve for the relaxation rate we compute the two eigenvalues:

\begin{align*}
\mu_{1,2} & =\frac{-\frac{\gamma}{2}\pm\sqrt{\left(\frac{\gamma}{2}\right)^{2}-\lambda\gamma}}{2}.
\end{align*}
When, $\gamma-4\lambda>0$ convergence is exponential. When $\gamma-4\lambda<0$
there is also an oscillatory contribution. To ensure that our result
is consistent with the case of full relaxation we take $\lambda/\gamma$
to be small, and expand the eigenvalues in a Taylor series: 

\begin{align*}
\mu_{1} & \approx-\frac{\gamma}{2}\\
\mu_{2} & \approx-\frac{\lambda}{2}
\end{align*}
Since convergence depends on the smaller eigenvalue, $\lambda$, our
results are consistent with the case of complete relaxation.

\bibliographystyle{apsrmp4-2}
\bibliography{biblo}

\begin{thebibliography}{23}%
\makeatletter
\providecommand \@ifxundefined [1]{%
 \@ifx{#1\undefined}
}%
\providecommand \@ifnum [1]{%
 \ifnum #1\expandafter \@firstoftwo
 \else \expandafter \@secondoftwo
 \fi
}%
\providecommand \@ifx [1]{%
 \ifx #1\expandafter \@firstoftwo
 \else \expandafter \@secondoftwo
 \fi
}%
\providecommand \natexlab [1]{#1}%
\providecommand \enquote  [1]{``#1''}%
\providecommand \bibnamefont  [1]{#1}%
\providecommand \bibfnamefont [1]{#1}%
\providecommand \citenamefont [1]{#1}%
\providecommand \href@noop [0]{\@secondoftwo}%
\providecommand \href [0]{\begingroup \@sanitize@url \@href}%
\providecommand \@href[1]{\@@startlink{#1}\@@href}%
\providecommand \@@href[1]{\endgroup#1\@@endlink}%
\providecommand \@sanitize@url [0]{\catcode `\\12\catcode `\$12\catcode
  `\&12\catcode `\#12\catcode `\^12\catcode `\_12\catcode `\%12\relax}%
\providecommand \@@startlink[1]{}%
\providecommand \@@endlink[0]{}%
\providecommand \url  [0]{\begingroup\@sanitize@url \@url }%
\providecommand \@url [1]{\endgroup\@href {#1}{\urlprefix }}%
\providecommand \urlprefix  [0]{URL }%
\providecommand \Eprint [0]{\href }%
\providecommand \doibase [0]{https://doi.org/}%
\providecommand \selectlanguage [0]{\@gobble}%
\providecommand \bibinfo  [0]{\@secondoftwo}%
\providecommand \bibfield  [0]{\@secondoftwo}%
\providecommand \translation [1]{[#1]}%
\providecommand \BibitemOpen [0]{}%
\providecommand \bibitemStop [0]{}%
\providecommand \bibitemNoStop [0]{.\EOS\space}%
\providecommand \EOS [0]{\spacefactor3000\relax}%
\providecommand \BibitemShut  [1]{\csname bibitem#1\endcsname}%
\let\auto@bib@innerbib\@empty
\bibitem [{\citenamefont {Anisetti}\ \emph {et~al.}(2022)\citenamefont
  {Anisetti}, \citenamefont {Scellier},\ and\ \citenamefont
  {Schwarz}}]{anisetti2022learning}%
  \BibitemOpen
  \bibfield  {author} {\bibinfo {author} {\bibnamefont {Anisetti},
  \bibfnamefont {V.~R.}}, \bibinfo {author} {\bibfnamefont {B.}~\bibnamefont
  {Scellier}}, and\ \bibinfo {author} {\bibfnamefont {J.}~\bibnamefont
  {Schwarz}}} (\bibinfo {year} {2022}),\ \href@noop {} {\bibinfo  {journal}
  {arXiv preprint arXiv:2203.12098}\ }\BibitemShut {NoStop}%
\bibitem [{\citenamefont {Bhattacharyya}\ \emph {et~al.}(2022)\citenamefont
  {Bhattacharyya}, \citenamefont {Zwicker},\ and\ \citenamefont
  {Alim}}]{bhattacharyya2022memory}%
  \BibitemOpen
\bibfield  {journal} {  }\bibfield  {author} {\bibinfo {author} {\bibnamefont
  {Bhattacharyya}, \bibfnamefont {K.}}, \bibinfo {author} {\bibfnamefont
  {D.}~\bibnamefont {Zwicker}}, and\ \bibinfo {author} {\bibfnamefont
  {K.}~\bibnamefont {Alim}}} (\bibinfo {year} {2022}),\ \href@noop {}
  {\bibfield  {journal} {\bibinfo  {journal} {Physical Review Letters}\
  }\textbf {\bibinfo {volume} {129}}~(\bibinfo {number} {2}),\ \bibinfo {pages}
  {028101}}\BibitemShut {NoStop}%
\bibitem [{\citenamefont {Calladine}(1978)}]{CALLADINE}%
  \BibitemOpen
  \bibfield  {author} {\bibinfo {author} {\bibnamefont {Calladine},
  \bibfnamefont {C.}}} (\bibinfo {year} {1978}),\ \href@noop {} {\bibfield
  {journal} {\bibinfo  {journal} {International Journal of Solids and
  Structures}\ }\textbf {\bibinfo {volume} {14}},\ \bibinfo {pages}
  {161}}\BibitemShut {NoStop}%
\bibitem [{\citenamefont {Dillavou}\ \emph {et~al.}(2022)\citenamefont
  {Dillavou}, \citenamefont {Stern}, \citenamefont {Liu},\ and\ \citenamefont
  {Durian}}]{dillavou2022demonstration}%
  \BibitemOpen
  \bibfield  {author} {\bibinfo {author} {\bibnamefont {Dillavou},
  \bibfnamefont {S.}}, \bibinfo {author} {\bibfnamefont {M.}~\bibnamefont
  {Stern}}, \bibinfo {author} {\bibfnamefont {A.~J.}\ \bibnamefont {Liu}}, and\
  \bibinfo {author} {\bibfnamefont {D.~J.}\ \bibnamefont {Durian}}} (\bibinfo
  {year} {2022}),\ \href@noop {} {\bibfield  {journal} {\bibinfo  {journal}
  {Physical Review Applied}\ }\textbf {\bibinfo {volume} {18}}~(\bibinfo
  {number} {1}),\ \bibinfo {pages} {014040}}\BibitemShut {NoStop}%
\bibitem [{\citenamefont {Goodrich}\ \emph {et~al.}(2015)\citenamefont
  {Goodrich}, \citenamefont {Liu},\ and\ \citenamefont
  {Nagel}}]{goodrich2015principle}%
  \BibitemOpen
  \bibfield  {author} {\bibinfo {author} {\bibnamefont {Goodrich},
  \bibfnamefont {C.~P.}}, \bibinfo {author} {\bibfnamefont {A.~J.}\
  \bibnamefont {Liu}}, and\ \bibinfo {author} {\bibfnamefont {S.~R.}\
  \bibnamefont {Nagel}}} (\bibinfo {year} {2015}),\ \href@noop {} {\bibfield
  {journal} {\bibinfo  {journal} {Physical review letters}\ }\textbf {\bibinfo
  {volume} {114}}~(\bibinfo {number} {22}),\ \bibinfo {pages}
  {225501}}\BibitemShut {NoStop}%
\bibitem [{\citenamefont {Hagh}\ \emph {et~al.}(2022)\citenamefont {Hagh},
  \citenamefont {Nagel}, \citenamefont {Liu}, \citenamefont {Manning},\ and\
  \citenamefont {Corwin}}]{hagh2022transient}%
  \BibitemOpen
  \bibfield  {author} {\bibinfo {author} {\bibnamefont {Hagh}, \bibfnamefont
  {V.~F.}}, \bibinfo {author} {\bibfnamefont {S.~R.}\ \bibnamefont {Nagel}},
  \bibinfo {author} {\bibfnamefont {A.~J.}\ \bibnamefont {Liu}}, \bibinfo
  {author} {\bibfnamefont {M.~L.}\ \bibnamefont {Manning}}, and\ \bibinfo
  {author} {\bibfnamefont {E.~I.}\ \bibnamefont {Corwin}}} (\bibinfo {year}
  {2022}),\ \href@noop {} {\bibfield  {journal} {\bibinfo  {journal}
  {Proceedings of the National Academy of Sciences}\ }\textbf {\bibinfo
  {volume} {119}}~(\bibinfo {number} {19}),\ \bibinfo {pages}
  {e2117622119}}\BibitemShut {NoStop}%
\bibitem [{\citenamefont {Hexner}\ \emph {et~al.}(2020)\citenamefont {Hexner},
  \citenamefont {Liu},\ and\ \citenamefont {Nagel}}]{hexner2020periodic}%
  \BibitemOpen
  \bibfield  {author} {\bibinfo {author} {\bibnamefont {Hexner}, \bibfnamefont
  {D.}}, \bibinfo {author} {\bibfnamefont {A.~J.}\ \bibnamefont {Liu}}, and\
  \bibinfo {author} {\bibfnamefont {S.~R.}\ \bibnamefont {Nagel}}} (\bibinfo
  {year} {2020}),\ \href@noop {} {\bibfield  {journal} {\bibinfo  {journal}
  {Proceedings of the National Academy of Sciences}\ }\textbf {\bibinfo
  {volume} {117}}~(\bibinfo {number} {50}),\ \bibinfo {pages}
  {31690}}\BibitemShut {NoStop}%
\bibitem [{\citenamefont {Kendall}\ \emph {et~al.}(2020)\citenamefont
  {Kendall}, \citenamefont {Pantone}, \citenamefont {Manickavasagam},
  \citenamefont {Bengio},\ and\ \citenamefont
  {Scellier}}]{kendall2020training}%
  \BibitemOpen
  \bibfield  {author} {\bibinfo {author} {\bibnamefont {Kendall}, \bibfnamefont
  {J.}}, \bibinfo {author} {\bibfnamefont {R.}~\bibnamefont {Pantone}},
  \bibinfo {author} {\bibfnamefont {K.}~\bibnamefont {Manickavasagam}},
  \bibinfo {author} {\bibfnamefont {Y.}~\bibnamefont {Bengio}}, and\ \bibinfo
  {author} {\bibfnamefont {B.}~\bibnamefont {Scellier}}} (\bibinfo {year}
  {2020}),\ \href@noop {} {\bibinfo  {journal} {arXiv preprint
  arXiv:2006.01981}\ }\BibitemShut {NoStop}%
\bibitem [{\citenamefont {Lubensky}\ \emph {et~al.}(2015)\citenamefont
  {Lubensky}, \citenamefont {Kane}, \citenamefont {Mao}, \citenamefont
  {Souslov},\ and\ \citenamefont {Sun}}]{Lubensky_rev}%
  \BibitemOpen
\bibfield  {journal} {  }\bibfield  {author} {\bibinfo {author} {\bibnamefont
  {Lubensky}, \bibfnamefont {T.~C.}}, \bibinfo {author} {\bibfnamefont {C.~L.}\
  \bibnamefont {Kane}}, \bibinfo {author} {\bibfnamefont {X.}~\bibnamefont
  {Mao}}, \bibinfo {author} {\bibfnamefont {A.}~\bibnamefont {Souslov}}, and\
  \bibinfo {author} {\bibfnamefont {K.}~\bibnamefont {Sun}}} (\bibinfo {year}
  {2015}),\ \href@noop {} {\bibfield  {journal} {\bibinfo  {journal} {Reports
  on Progress in Physics}\ }\textbf {\bibinfo {volume} {78}}~(\bibinfo {number}
  {7}),\ \bibinfo {pages} {073901}}\BibitemShut {NoStop}%
\bibitem [{\citenamefont {Maxwell}(1864)}]{maxwell1864calculation}%
  \BibitemOpen
  \bibfield  {author} {\bibinfo {author} {\bibnamefont {Maxwell}, \bibfnamefont
  {J.~C.}}} (\bibinfo {year} {1864}),\ \href@noop {} {\bibfield  {journal}
  {\bibinfo  {journal} {The London, Edinburgh, and Dublin Philosophical
  Magazine and Journal of Science}\ }\textbf {\bibinfo {volume} {27}}~(\bibinfo
  {number} {182}),\ \bibinfo {pages} {294}}\BibitemShut {NoStop}%
\bibitem [{Note1(????)}]{Note1}%
  \BibitemOpen
  \bibinfo {note} {Previous studies on training found that the particular
  ensemble is unimportant.}\BibitemShut {Stop}%
\bibitem [{Note2(????)}]{Note2}%
  \BibitemOpen
  \bibinfo {note} {In the appendix we compute the convergence rate for a simple
  case.}\BibitemShut {Stop}%
\bibitem [{\citenamefont {O'Hern}\ \emph {et~al.}(2003)\citenamefont {O'Hern},
  \citenamefont {Silbert}, \citenamefont {Liu},\ and\ \citenamefont
  {Nagel}}]{Ohern}%
  \BibitemOpen
  \bibfield  {author} {\bibinfo {author} {\bibnamefont {O'Hern}, \bibfnamefont
  {C.~S.}}, \bibinfo {author} {\bibfnamefont {L.~E.}\ \bibnamefont {Silbert}},
  \bibinfo {author} {\bibfnamefont {A.~J.}\ \bibnamefont {Liu}}, and\ \bibinfo
  {author} {\bibfnamefont {S.~R.}\ \bibnamefont {Nagel}}} (\bibinfo {year}
  {2003}),\ \href@noop {} {\bibfield  {journal} {\bibinfo  {journal} {Phys.
  Rev. E}\ }\textbf {\bibinfo {volume} {68}},\ \bibinfo {pages}
  {011306}}\BibitemShut {NoStop}%
\bibitem [{\citenamefont {Pashine}\ \emph {et~al.}(2019)\citenamefont
  {Pashine}, \citenamefont {Hexner}, \citenamefont {Liu},\ and\ \citenamefont
  {Nagel}}]{pashine2019directed}%
  \BibitemOpen
  \bibfield  {author} {\bibinfo {author} {\bibnamefont {Pashine}, \bibfnamefont
  {N.}}, \bibinfo {author} {\bibfnamefont {D.}~\bibnamefont {Hexner}}, \bibinfo
  {author} {\bibfnamefont {A.~J.}\ \bibnamefont {Liu}}, and\ \bibinfo {author}
  {\bibfnamefont {S.~R.}\ \bibnamefont {Nagel}}} (\bibinfo {year} {2019}),\
  \href@noop {} {\bibfield  {journal} {\bibinfo  {journal} {Science advances}\
  }\textbf {\bibinfo {volume} {5}}~(\bibinfo {number} {12}),\ \bibinfo {pages}
  {eaax4215}}\BibitemShut {NoStop}%
\bibitem [{\citenamefont {Pisanty}\ \emph {et~al.}(2021)\citenamefont
  {Pisanty}, \citenamefont {O{\u{g}}uz}, \citenamefont {Nisoli},\ and\
  \citenamefont {Shokef}}]{pisanty2021putting}%
  \BibitemOpen
  \bibfield  {author} {\bibinfo {author} {\bibnamefont {Pisanty}, \bibfnamefont
  {B.}}, \bibinfo {author} {\bibfnamefont {E.~C.}\ \bibnamefont {O{\u{g}}uz}},
  \bibinfo {author} {\bibfnamefont {C.}~\bibnamefont {Nisoli}}, and\ \bibinfo
  {author} {\bibfnamefont {Y.}~\bibnamefont {Shokef}}} (\bibinfo {year}
  {2021}),\ \href@noop {} {\bibfield  {journal} {\bibinfo  {journal} {SciPost
  Physics}\ }\textbf {\bibinfo {volume} {10}}~(\bibinfo {number} {6}),\
  \bibinfo {pages} {136}}\BibitemShut {NoStop}%
\bibitem [{\citenamefont {Rocks}\ \emph {et~al.}(2019)\citenamefont {Rocks},
  \citenamefont {Ronellenfitsch}, \citenamefont {Liu}, \citenamefont {Nagel},\
  and\ \citenamefont {Katifori}}]{rocks2019limits}%
  \BibitemOpen
  \bibfield  {author} {\bibinfo {author} {\bibnamefont {Rocks}, \bibfnamefont
  {J.~W.}}, \bibinfo {author} {\bibfnamefont {H.}~\bibnamefont
  {Ronellenfitsch}}, \bibinfo {author} {\bibfnamefont {A.~J.}\ \bibnamefont
  {Liu}}, \bibinfo {author} {\bibfnamefont {S.~R.}\ \bibnamefont {Nagel}}, and\
  \bibinfo {author} {\bibfnamefont {E.}~\bibnamefont {Katifori}}} (\bibinfo
  {year} {2019}),\ \href@noop {} {\bibfield  {journal} {\bibinfo  {journal}
  {Proceedings of the National Academy of Sciences}\ }\textbf {\bibinfo
  {volume} {116}}~(\bibinfo {number} {7}),\ \bibinfo {pages}
  {2506}}\BibitemShut {NoStop}%
\bibitem [{\citenamefont {Sartor}\ and\ \citenamefont
  {Corwin}(2022)}]{sartor2022predicting}%
  \BibitemOpen
  \bibfield  {author} {\bibinfo {author} {\bibnamefont {Sartor}, \bibfnamefont
  {J.~D.}}, and\ \bibinfo {author} {\bibfnamefont {E.~I.}\ \bibnamefont
  {Corwin}}} (\bibinfo {year} {2022}),\ \href@noop {} {\bibfield  {journal}
  {\bibinfo  {journal} {Physical Review Letters}\ }\textbf {\bibinfo {volume}
  {129}}~(\bibinfo {number} {18}),\ \bibinfo {pages} {188001}}\BibitemShut
  {NoStop}%
\bibitem [{\citenamefont {Scellier}\ and\ \citenamefont
  {Bengio}(2017)}]{scellier2017equilibrium}%
  \BibitemOpen
  \bibfield  {author} {\bibinfo {author} {\bibnamefont {Scellier},
  \bibfnamefont {B.}}, and\ \bibinfo {author} {\bibfnamefont {Y.}~\bibnamefont
  {Bengio}}} (\bibinfo {year} {2017}),\ \href@noop {} {\bibfield  {journal}
  {\bibinfo  {journal} {Frontiers in computational neuroscience}\ }\textbf
  {\bibinfo {volume} {11}},\ \bibinfo {pages} {24}}\BibitemShut {NoStop}%
\bibitem [{\citenamefont {Stern}\ \emph {et~al.}(2020)\citenamefont {Stern},
  \citenamefont {Arinze}, \citenamefont {Perez}, \citenamefont {Palmer},\ and\
  \citenamefont {Murugan}}]{stern2020supervised2}%
  \BibitemOpen
  \bibfield  {author} {\bibinfo {author} {\bibnamefont {Stern}, \bibfnamefont
  {M.}}, \bibinfo {author} {\bibfnamefont {C.}~\bibnamefont {Arinze}}, \bibinfo
  {author} {\bibfnamefont {L.}~\bibnamefont {Perez}}, \bibinfo {author}
  {\bibfnamefont {S.~E.}\ \bibnamefont {Palmer}}, and\ \bibinfo {author}
  {\bibfnamefont {A.}~\bibnamefont {Murugan}}} (\bibinfo {year} {2020}),\
  \href@noop {} {\bibfield  {journal} {\bibinfo  {journal} {Proceedings of the
  National Academy of Sciences}\ }\textbf {\bibinfo {volume} {117}}~(\bibinfo
  {number} {26}),\ \bibinfo {pages} {14843}}\BibitemShut {NoStop}%
\bibitem [{\citenamefont {Stern}\ \emph {et~al.}(2022)\citenamefont {Stern},
  \citenamefont {Dillavou}, \citenamefont {Miskin}, \citenamefont {Durian},\
  and\ \citenamefont {Liu}}]{stern2022physical}%
  \BibitemOpen
  \bibfield  {author} {\bibinfo {author} {\bibnamefont {Stern}, \bibfnamefont
  {M.}}, \bibinfo {author} {\bibfnamefont {S.}~\bibnamefont {Dillavou}},
  \bibinfo {author} {\bibfnamefont {M.~Z.}\ \bibnamefont {Miskin}}, \bibinfo
  {author} {\bibfnamefont {D.~J.}\ \bibnamefont {Durian}}, and\ \bibinfo
  {author} {\bibfnamefont {A.~J.}\ \bibnamefont {Liu}}} (\bibinfo {year}
  {2022}),\ \href@noop {} {\bibfield  {journal} {\bibinfo  {journal} {Physical
  Review Research}\ }\textbf {\bibinfo {volume} {4}}~(\bibinfo {number} {2}),\
  \bibinfo {pages} {L022037}}\BibitemShut {NoStop}%
\bibitem [{\citenamefont {Stern}\ \emph {et~al.}(2021)\citenamefont {Stern},
  \citenamefont {Hexner}, \citenamefont {Rocks},\ and\ \citenamefont
  {Liu}}]{stern2021supervised}%
  \BibitemOpen
  \bibfield  {author} {\bibinfo {author} {\bibnamefont {Stern}, \bibfnamefont
  {M.}}, \bibinfo {author} {\bibfnamefont {D.}~\bibnamefont {Hexner}}, \bibinfo
  {author} {\bibfnamefont {J.~W.}\ \bibnamefont {Rocks}}, and\ \bibinfo
  {author} {\bibfnamefont {A.~J.}\ \bibnamefont {Liu}}} (\bibinfo {year}
  {2021}),\ \href@noop {} {\bibfield  {journal} {\bibinfo  {journal} {Physical
  Review X}\ }\textbf {\bibinfo {volume} {11}}~(\bibinfo {number} {2}),\
  \bibinfo {pages} {021045}}\BibitemShut {NoStop}%
\bibitem [{\citenamefont {Vojta}(2006)}]{vojta2006rare}%
  \BibitemOpen
  \bibfield  {author} {\bibinfo {author} {\bibnamefont {Vojta}, \bibfnamefont
  {T.}}} (\bibinfo {year} {2006}),\ \href@noop {} {\bibfield  {journal}
  {\bibinfo  {journal} {Journal of Physics A: Mathematical and General}\
  }\textbf {\bibinfo {volume} {39}}~(\bibinfo {number} {22}),\ \bibinfo {pages}
  {R143}}\BibitemShut {NoStop}%
\bibitem [{\citenamefont {Whitesides}\ and\ \citenamefont
  {Grzybowski}(2002)}]{whitesides2002self}%
  \BibitemOpen
  \bibfield  {author} {\bibinfo {author} {\bibnamefont {Whitesides},
  \bibfnamefont {G.~M.}}, and\ \bibinfo {author} {\bibfnamefont
  {B.}~\bibnamefont {Grzybowski}}} (\bibinfo {year} {2002}),\ \href@noop {}
  {\bibfield  {journal} {\bibinfo  {journal} {Science}\ }\textbf {\bibinfo
  {volume} {295}}~(\bibinfo {number} {5564}),\ \bibinfo {pages}
  {2418}}\BibitemShut {NoStop}%
\end{thebibliography}%

\end{document}